\documentclass[12pt]{article}
\usepackage{times,epsfig,cite,amssymb}
\usepackage{graphicx}

\normalbaselines
\title{Hyperconfluent third-order \\
supersymmetric quantum mechanics}

\author{David J. Fern\'andez C.$^\dag$ \\
Encarnaci\'on Salinas-Hern\'andez$^\S$
\\
[12pt] {\small
$^\dag$~Departamento de F\'isica, Cinvestav} \\ 
{\small A.P. 14-740, 07000 M\'exico DF, Mexico} 
\\ [12pt] 
{\small $^\S$~Escuela Superior de C\'omputo, 
Instituto Polit\'ecnico Nacional} \\ 
{\small Ed. 15, U.P. Adolfo L\'opez Mateos, 
07738 M\'exico D.F., Mexico}}

\date{}

\begin{document}

\maketitle

\begin{abstract}
The hyperconfluent third-order supersymmetric quantum mechanics, in which all
the factorization energies tend to a common value, is analyzed.
It will be shown that the final potential as well can be achieved by applying
consecutively a confluent second-order and a first-order SUSY transformations, both
with the same factorization energy. The technique will be applied to the free particle
and the Coulomb potential.
\end{abstract}


\section{Introduction}

Nowadays, {\it Supersymmetric Quantum Mechanics} (SUSY QM) has became the standard technique
for generating solvable potentials from a given initial one \cite{cks95,ju96,ba01,mr04,ff05,fe10}.
Moreover, it represents a powerful tool for designing Hamiltonians with a fixed prescribed
spectrum (see, e.g, \cite{as03,ac04,fgn11}). In its higher-order version, in which the differential
intertwining operators are of order greater that one, it is well known that several seed
solutions of the initial stationary Sch\"odinger equation with an appropriate behavior are in
general required for calculating the new potential, the eigenfunctions of the corresponding
Hamiltonian, etc. \cite{ais93,aicd95,sa96,fe97,fhm98,ro98a,ro98b,anst01,cjnp08}. If for some
reason just one of those seeds is available, one is driven to what could be called as {\it
hyperconfluent higher-order SUSY QM}, in which all the factorization energies which are involved
tend to a common value.

In the past several works dealing with the {\it confluent second-order SUSY QM} have been
elaborated \cite{mnr00,fs03,fs05,bu10}. Up to our knowledge, however,
the {\it hyperconfluent third-order SUSY QM} has not been addressed explicitly. Of course, several
papers involving the third-order SUSY QM have been done, but they are centered mainly in the case
when the factorization energies are all different (see e.g. \cite{acin00,fnn04,cfnn04,in04,mn08,
rr08,bf11} and references therein).

In this article we aim to fill the gap by studying in detail the hyperconfluent third-order
SUSY QM. In order to achieve this, we have arranged the paper as follows. In the next Section
we will briefly review the confluent second-order SUSY QM. In Section 3 we will
analyze the direct approach to the hyperconfluent third-order SUSY QM, while
in Section 4 we will address the corresponding iterative method. Section 5 explores the
requirements that the seed solution has to obey in order to produce non-singular transformations
as well as the eigenfunctions of the SUSY generated Hamiltonians.
In Section 6 we will illustrate our general treatment by means of two specific examples, the free
particle and the Coulomb potential. Our conclusions will be presented in Section 7.

\section{Confluent second-order SUSY QM}

Let us consider a one-dimensional Schr\"odinger Hamiltonian
\begin{eqnarray}\label{hzero}
& H_0 = - \frac{d^2}{dx^2} + V_0(x).
\end{eqnarray}
The domain of definition or the corresponding system is denoted as ${\cal D} = [x_l,x_r]$.
Thus, depending on the problem we are dealing with, and the consequent identification of
$x_l$ and $x_r$, this domain could be the full real line, the positive semi-axis or a finite
interval. The eigenfunctions and eigenvalues associated to the discrete part of the spectrum
of $H_0$, denoted by $\psi_n(x), \ E_n, \ n=0,1,\dots$, satisfy the stationary Schr\"odinger
equation
\begin{eqnarray}
& H_0 \psi_n = - \psi_n'' + V_0 \psi_n = E_n \psi_n,
\end{eqnarray}
as well as the boundary conditions
\begin{eqnarray}\label{bondaryconditions}
& \psi_n(x_l) = \psi_n(x_r) = 0.
\end{eqnarray}

From now on we are going to suppose that all the eigenfunctions and eigenvalues of $H_0$ are
known. In the general formulation of the second-order SUSY QM one looks for a new Hamiltonian
\begin{eqnarray}\label{htwo}
& H_2 = - \frac{d^2}{dx^2} + V_2(x),
\end{eqnarray}
which is intertwined with $H_0$ by a second-order operator $B_2^+$ in the way
\begin{eqnarray}\label{intertwining2}
& H_2B_2^+ = B_2^+ H_0,
\end{eqnarray}
where
\begin{eqnarray}\label{btwo}
& B_2^+ = \frac{d^2}{dx^2} - \eta(x) \frac{d}{dx} + \gamma(x).
\end{eqnarray}
By plugging these expressions in the intertwining relationship (\ref{intertwining2}),
decoupling the resulting system of equations and solving it we arrive at \cite{ff05,fe10}:
\begin{eqnarray}
V_2 & = & V_0 - 2\eta', \\
\gamma & = & \frac{\eta'}2 + \frac{\eta^2}2 - V_0 + \frac{\epsilon_1 + \epsilon_2}2, \\
\eta & = & \{\ln[W(u_1,u_2)]\}',
\end{eqnarray}
where $u_1, \ u_2$ are two seed solutions of the initial stationary Schr\"odinger equation
associated to the factorization energies $\epsilon_1, \ \epsilon_2$ (in general different)
\begin{eqnarray}\label{eqseeds}
& H_0 u_i = - u_i'' + V_0 u_i = \epsilon_i u_i, \qquad i=1,2,
\end{eqnarray}
and $W(u_1,u_2) = u_1 u_2' - u_1' u_2$ denotes their Wronskian. Note that the seeds
$u_1$, $u_2$ could obey or not the boundary conditions of equation
(\ref{bondaryconditions}).

The {\it confluent second-order SUSY QM} arises now as a limit procedure of the previous formalism when
$\epsilon_1\rightarrow \epsilon_2 \rightarrow \epsilon$ \cite{mnr00}. Note that, if the potential $V_2$
is going to be different from the initial one, then $u_1$ and $u_2$ cannot be just chosen as two linearly
independent solutions of equation (\ref{eqseeds}), since then $W(u_1,u_2) = {\rm constant}$ and therefore
$V_2 = V_0$. In order to produce non-trivial results, the right choice is to take $u_1$ as a standard
eigenfunction of $H_0$ while $u_2$ becomes a generalized eigenfunction of rank $2$ of $H_0$, both associated
to $\epsilon$, namely:
\begin{eqnarray}
&& \hskip-1cm (H_0 - \epsilon) u_1 = 0 \quad \Rightarrow \quad u_1'' = (V_0 - \epsilon) u_1 , \label{difu1} \\
&& \hskip-1cm (H_0 - \epsilon) u_2 = u_1 \quad \Rightarrow \quad (H_0 - \epsilon)^2 u_2 = 0
\quad \Rightarrow \quad  u_2'' = (V_0 - \epsilon) u_2 - u_1, \label{difu2}
\end{eqnarray}
i.e., we are employing a Jordan chain of length $2$. Expressing this in matrix language \cite{as03,ac04},
this specific choice of basis $\{u_1, u_2\}$ means that, in the restriction to the two-dimensional
subspace of functions belonging to Ker$(B_2^+)$, the initial Hamiltonian $H_0$ is represented by a matrix
$(H_0)$ having a non-trivial Jordan structure of standard type:
\begin{eqnarray}\label{jchl2}
(H_0) = \left(
\matrix{
 \epsilon & 1 \cr 0 & \epsilon}
 \right).
\end{eqnarray}

Note that, if $u_1$ is given, then it is possible to determine
the general solution $u_2$ to the second-order equation (\ref{difu2}) and to calculate then explicitly
$W(u_1,u_2)$ \cite{fs05}. An alternative (and shorter) procedure runs as follows. First of all it is
straightforward to show that
\begin{equation}
W'(u_1,u_2) = u_1 u_2'' - u_2 u_1'' = - u_1^2,
\end{equation}
where we have used equations (\ref{difu1}, \ref{difu2}). This implies that:
\begin{equation} \label{2conf}
W(u_1,u_2) = w_0 - \int_{x_0}^x u_1^2(y) dy \equiv w(x),
\end{equation}
where $x_0 \in (x_l,x_r)$.

Let us note that, in order that the new potential
\begin{eqnarray}
V_2 & = & V_0 - 2\{\ln[W(u_1,u_2)]\}'' = V_0 - 2[\ln(w)]''
\end{eqnarray}
has not additional singularities with respect to $V_0$, then  $w(x)$ must not have nodes in
$(x_l,x_r)$. This can be achieved by choosing a Schr\"odinger seed solution $u_1$ such that
\cite{fs03}:
\begin{eqnarray} \label{uplus}
&& \lim\limits_{x\rightarrow x_l} u_1 = 0, \qquad
\nu_- \equiv  \int_{x_l}^{x_0} u_1^2(y) \, dy <\infty, \qquad {\rm or} \\
&& \lim\limits_{x\rightarrow x_r} u_1 = 0, \qquad
\nu_+ \equiv \int_{x_0}^{x_r} u_1^2(y) \, dy <\infty.
\end{eqnarray}
With this choice, it turns out that $w(x)$ becomes nodeless either for $w_0 \leq - \nu_-$ in
the first case or for $w_0 \geq \nu_+$ in the second one. Moreover, departing from the
normalized bound states $\psi_n(x)$ of $H_0$ the normalized ones $\psi_n^{(2)}(x)$ of $H_2$
can be built up in the way:
\begin{eqnarray}
&&  \psi_n^{(2)}(x) = \frac{B_2^+ \psi_n(x)}{E_n - \epsilon}.
\end{eqnarray}
In addition, there is an eigenfunction of $H_2$ associated with $\epsilon$ which becomes as well
square integrable (we are using here a notation for this state which is appropriate for the
purpose of this paper):
\begin{eqnarray}\label{missingtwo}
&&  u_1^{(2)}(x) \propto \frac{u_1(x)}{w(x)}.
\end{eqnarray}

Note that the confluent algorithm has been used to create bound states above the ground state
energy of $H_0$ \cite{fs03}. This possibility of spectral manipulation typically was outside
the goals of the standard first-order SUSY QM. Moreover, the use of just one eigenfunction of $H_0$
in the confluent case is advantageous compared with the  second-order SUSY QM with $\epsilon_1
\neq\epsilon_2$, which requires the knowledge of two Schr\"odinger seed solutions.

\section{Hyperconfluent third-order SUSY QM: direct approach}

In turn, let us analyze the hyperconfluent third-order SUSY QM, for which the three factorization
energies converge to the same $\epsilon$-value, namely $\epsilon_i \rightarrow \epsilon, \
i=1,2,3$. Similarly as for the second-order case of Section 2, we are going to use here a Jordan
chain of length $3$ of generalized eigenfunctions $\{u_1,u_2,u_3\}$ such that $u_1, \ u_2$ obey equations
(\ref{difu1},\ref{difu2}) while $u_3$ satisfies
\begin{eqnarray}
&& \hskip-1cm (H_0 - \epsilon) u_3 = u_2 \quad \Rightarrow \quad (H_0 - \epsilon)^3 u_3 = 0 \quad
\Rightarrow \quad  u_3'' = (V_0 - \epsilon) u_3 - u_2. \label{difu3}
\end{eqnarray}
Equations (\ref{difu1},\ref{difu2},\ref{difu3}) mean that in the
three-dimensional subspace of functions belonging to Ker$(B_3^+)$ this choice of basis implies that the
matrix representing to $H_0$ has a non-trivial Jordan structure of standard type
\footnote[1]{Note that the matrices $(H_0)$ of equations (\ref{jchl2}) and (\ref{jchl3}) are non-hermitian
despite $H_0$ is hermitian.}
\begin{eqnarray}\label{jchl3}
(H_0) = \left(
\matrix{
 \epsilon & 1 & 0 \cr 0 & \epsilon & 1 \cr 0 & 0 & \epsilon}
 \right).
\end{eqnarray}

Now, the hyperconfluent third-order SUSY partner Hamiltonians $H_0$ and $H_3$
are intertwined by the third-order operator $B_3^+$ in the way
\begin{eqnarray}\label{intertwining3}
& H_3 B_3^+ = B_3^+ H_0,
\end{eqnarray}
where $H_0$ is given by equation (\ref{hzero}), $H_3$ has the standard Schr\"odinger form
\begin{eqnarray}\label{hthree}
& H_3 = - \frac{d^2}{dx^2} + V_3(x),
\end{eqnarray}
and $V_3$ is expressed in terms of the initial potential and the three seeds $u_1, \ u_2, \ u_3$
in the way:
\begin{equation} \label{v3a}
V_3 = V_0 - 2 \{\ln [W(u_1, u_2, u_3)] \}'',
\end{equation}
with $W(u_1, u_2, u_3)$ denoting the Wronskian of $u_1$, $u_2$ and $u_3$ (we will give the explicit
expression for $B_3^+$ in the next Section). A straightforward calculation leads to:
\begin{eqnarray}
\hskip-1.5cm && W(u_1, u_2, u_3) =
\left\vert
\matrix{ u_1 & u_2 & u_3 \cr
u_1' & u_2' & u_3' \cr u_1'' & u_2'' & u_3''}
\right\vert = u_1'' W(u_2,u_3) - u_2'' W(u_1,u_3) +
u_3'' W(u_1,u_2) .
\end{eqnarray}
By using now equations (\ref{difu1},\ref{difu2},\ref{difu3}) it turns out that:
\begin{eqnarray} \label{w123}
W(u_1, u_2, u_3) & = & u_1 W(u_1,u_3) - u_2 W(u_1,u_2) .
\end{eqnarray}
Recall that $W(u_1,u_2)=w(x)$ was calculated in a simple way in the previous Section;
thus, a similar procedure to obtain $W(u_1,u_3)$ can be followed, leading to:
\begin{equation} \label{w13}
W(u_1,u_3) = w_1 - \int_{x_0}^x u_1(y) u_2(y) dy.
\end{equation}
Given $u_1$, and consequently the $w$ of equation (\ref{2conf}), it remains just
to express $u_2$ in terms of them. Let us note first of all that
\begin{equation}
w = W(u_1,u_2) = u_1^2\left(\frac{u_2}{u_1}\right)'.
\end{equation}
Henceforth
\begin{equation}
u_2 = u_1\left[\beta_1 + \int_{x_0}^x \frac{w(y)}{u_1^2(y)}dy\right].
\end{equation}
Thus, a straightforward calculation leads to
\begin{equation} \label{int12}
\hskip-1.5cm \int_{x_0}^x u_1(y) u_2(y) dy = w_0 \beta_1 - \beta_1 w(x) - w(x) \int_{x_0}^x \frac{w(y)}{u_1^2(y)}dy
+ \int_{x_0}^x \left[\frac{w(y)}{u_1(y)}\right]^2 dy.
\end{equation}
By plugging equations (\ref{2conf},\ref{w13},\ref{int12}) into equation
(\ref{w123}) we arrive at:
\begin{equation} \label{w123simplified}
W(u_1, u_2, u_3) = u_1 \left\{f_0 - \int_{x_0}^x \left[\frac{w(y)}{u_1(y)}\right]^2 dy\right\}
\equiv u_1 f,
\end{equation}
with
\begin{equation} \label{f(x)}
f(x) = f_0 - \int_{x_0}^x \left[\frac{w(y)}{u_1(y)}\right]^2 dy ,
\end{equation}
and $f_0 = w_1 - w_0 \beta_1$. Finally, the potential of equation (\ref{v3a})
becomes:
\begin{equation} \label{v3b}
V_3(x) = V_0(x) - 2 \{\ln [u_1(x)] \}''  - 2 \{\ln [f(x)]\}'',
\end{equation}
where $f(x)$ is given by equation (\ref{f(x)}).

\section{Hyperconfluent third-order SUSY QM: iterative approach}

We are going to apply now two consecutive SUSY transformations departing
from the initial Hamiltonian $H_0$: a confluent second-order one
for generating $V_2$ from $V_0$, which employs the two generalized
eigenfunctions $u_1, \ u_2$ associated to $\epsilon$ satisfying equations
(\ref{difu1},\ref{difu2}) of Section 2; then a first-order transformation
in order to obtain $V_3$ from $V_2$, which makes use of the general solution of the
stationary Schr\"odinger equation of $H_2$ associated to $\epsilon$.

As for the confluent second-order transformation, we saw at Section 2 that the new
potential $V_2$ is given by
\begin{equation}
V_2 = V_0 - 2 \{\ln [W(u_1,u_2)] \}'' = V_0 - 2 [\ln(w)]'' ,
\end{equation}
where the Wronskian $W(u_1,u_2) = w(x)$ of the two generalized eigenfunctions $u_1, \
u_2$ of $H_0$ associated to $\epsilon$ is given by equation (\ref{2conf}).

Concerning the first-order transformation, it turns out that $H_2$ and $H_3$ are intertwined
by a first-order operator $A_3^+$ in the way:
\begin{eqnarray}\label{intertwining1}
& H_3 A_3^+ = A_3^+ H_2,
\end{eqnarray}
where $H_2$ and $H_3$ are given by equations (\ref{htwo}) and (\ref{hthree}) respectively, and
\begin{eqnarray}\label{athree}
& A_3^+ = - \frac{d}{dx} + \ln[u^{(2)}]' = - \frac{d}{dx} + \frac{{u^{(2)}}'}{u^{(2)}},
\end{eqnarray}
with $u^{(2)}$ being the general solution of the Schr\"odinger equation
$$
H_2 u^{(2)} = \epsilon u^{(2)}.
$$
From the results of Section 2 it is known that one solution is given by $u^{(2)}_1=u_1/w$
(see equation (\ref{missingtwo}) and \cite{fs03,fs05}). The other linearly independent solution
$u^{(2)}_2$ is found by asking that $W(u^{(2)}_1,u^{(2)}_2)=1 =
[u^{(2)}_1]^2[u^{(2)}_2/u^{(2)}_1]'$, which immediately leads to $u^{(2)}_2 = u^{(2)}_1
\int_{x_0}^x dy/[u^{(2)}_1(y)]^2$. Thus, the solution $u^{(2)}$ we are looking for to implement
the first-order transformation takes the form:
\begin{equation} \label{usup2}
u^{(2)} = c_1 u^{(2)}_1 + c_2 u^{(2)}_2 = - c_2 \frac{u_1}{w}\left\{- \frac{c_1}{c_2} -
\int_{x_0}^x \left[\frac{w(y)}{u_1(y)}\right]^2dy\right\}.
\end{equation}
Hence, the final potential $V_3$ resulting from applying the first-order SUSY
transformation to the Hamiltonian $H_2$, when using the seed solution given in equation
(\ref{usup2}), becomes:
\begin{equation} \label{v3c}
V_3 = V_2 - 2 \{\ln [u^{(2)}] \}'' = V_0 - 2 \{\ln [u_1] \}'' - 2 \left\{\ln\left(- \frac{c_1}{c_2} -
\int_{x_0}^x \left[\frac{w(y)}{u_1(y)}\right]^2dy\right)\right\}''.
\end{equation}
Note that the two hyperconfluent third-order SUSY partner potentials $V_3(x)$ of $V_0(x)$ given by
equations (\ref{v3b}) and (\ref{v3c}) are exactly the same if it is taken $f_0 = - c_1/c_2$.

We can give, finally, the explicit expression for the third-order operator $B_3^+$ intertwining the
initial and final Hamiltonians $H_0$ and $H_3$ (see equation (\ref{intertwining3})):
\begin{equation} \label{bthree}
B_3^+ = A_3^+ B_2^+,
\end{equation}
where $B_2^+$ and $A_3^+$ are given by equations (\ref{btwo}) and (\ref{athree}) respectively.

\section{Non-singular transformations and bound states of $H_3$}

As can be seen from equation (\ref{v3a}), in order that the potential $V_3$ has no
additional singularities compared with those of $V_0$, the Wronskian $W(u_1,u_2,u_3)$
given in equation (\ref{w123simplified}) should not have nodes inside ${\cal D}$. This
implies that both functions $u_1$ and $f$ in this factorized expression should be free of
zeros in this domain, in particular the seed solution $u_1$ which automatically leads to
the restriction $\epsilon \leq E_0$, where $E_0$ is the ground state energy of $H_0$.
Moreover, for the second factor $f(x)$ of equation (\ref{f(x)}) to be nodeless,
the function $w/u_1$ should vanish either to the left edge $x_l$ of ${\cal D}$ or to the right
one $x_r$. Here we are going to discuss in detail just the first case; the second one can be
addressed in a similar way.

Let us choose first of all a nonphysical Schr\"odinger seed solution $u_1$ without nodes in
${\cal D}$, obeying equation (\ref{difu1}) for $\epsilon < E_0$. Moreover, it is supposed
that $u_1$ satisfies as well equation (\ref{uplus}).
Since $w/u_1$ should vanish for $x\rightarrow x_l$, we must have
\begin{equation}\label{limitxl0}
\lim_{x\rightarrow x_l} w = w_0 + \nu_- = 0,
\end{equation}
which implies that $w_0$ has to be taken as
\begin{equation}
w_0 = - \nu_- = - \int_{x_l}^{x_0} u_1^2(y) dy.
\end{equation}
Therefore:
\begin{equation} \label{wuplus}
w(x) = - \int_{x_l}^{x} u_1^2(y) dy.
\end{equation}
With this specific choice of $u_1$ and $w$, for most of the typical quantum mechanical
problems it turns out that:
\begin{equation}
\lim_{x\rightarrow x_l} \frac{w(x)}{u_1(x)} = 0 \quad {\rm and} \quad
\lim_{x\rightarrow x_r} \left\vert \frac{w(x)}{u_1(x)}\right\vert = \infty.
\end{equation}
Hence, the domain of the parameter $f_0$ such that $f(x)$ is nodeless in $(x_l,x_r)$ is given by
\begin{equation}\label{efecero}
f_0 < - \sigma_- = - \int_{x_l}^{x_0} \frac{w^2(y)}{u_1^2(y)} dy .
\end{equation}

Let us note that in this $f_0$-domain the third-order intertwining operator $B_3^+$ of equation
(\ref{bthree}) transforms the normalized eigenfunctions $\psi_n$ of $H_0$ into normalized
eigenfunctions $\psi_n^{(3)}$ of $H_3$ in the way:
\begin{equation}\label{boundsthree}
\psi_n^{(3)}(x) = \frac{B_3^+ \psi_n}{\sqrt{(E_n - \epsilon)^3}}.
\end{equation}
Moreover, the eigenfunction $\psi^{(3)}_{\epsilon}$ of $H_3$ associated to the eigenvalue
$\epsilon$ (compare equation (\ref{usup2})),
\begin{equation} \label{missings}
\psi^{(3)}_{\epsilon}(x) \propto \frac{1}{u^{(2)}(x)} \propto \frac{w(x)}{u_1(x)f(x)},
\end{equation}
turns out to be square-integrable in ${\cal D}$, which implies that the spectrum of $H_3$
becomes
\begin{equation}
{\rm Sp}(H_3) = \{\epsilon\}\cup {\rm Sp}(H_0) .
\end{equation}
Note that, when $f_0 \rightarrow - \sigma_-$, the hyperconfluent third-order
transformation remains non-singular but the eigenstate $\psi^{(3)}_{\epsilon}$ is not
longer square-integrable. Thus, in this limit the two Hamiltonians $H_3$ and $H_0$ become
isospectral.

On the other hand, for $u_1(x)$ being chosen as the normalized ground state
eigenfunction $\psi_0(x)$ of $H_0$ associated to $E_0$, it turns out that the previous equations
(\ref{limitxl0}-\ref{missings}) remain valid, the only difference is that now
$$
\nu_- = \int_{x_l}^{x_0} u_1^2(y) dy  < 1.
$$
Thus, for $f_0$ satisfying equation (\ref{efecero}), it turns out that $\epsilon = E_0
\in {\rm Sp}(H_3)$, which implies that
\begin{equation}
{\rm Sp}(H_3) = {\rm Sp}(H_0),
\end{equation}
i.e., the transformation is again strictly isospectral. However, when
$f_0 \rightarrow - \sigma_-$ the $\psi^{(3)}_{\epsilon}$ of equation
(\ref{missings}) is not longer normalizable, meaning that in this limit $\epsilon = E_0\not\in
{\rm Sp}(H_3)$, namely,
\begin{equation}
{\rm Sp}(H_3) = {\rm Sp}(H_0) - \{E_0\}.
\end{equation}
In this case, through the hyperconfluent third-order SUSY transformation
somehow we `delete' the ground state energy of $H_0$ for generating
$H_3$.

\section{Examples}

Let us apply next the previous formalism to two physically interesting
examples, the free particle and the Coulomb potential.

\subsection{Free particle}

The general solution of the stationary Schr\"odinger equation (\ref{difu1}) for
the free particle with a negative factorization energy $\epsilon = -k^2, \ k>0$
(for which $V_0(x) = 0$) is given by:
\begin{equation} \label{freegralu}
u_1(x) = A e^{kx} + B e^{-kx}.
\end{equation}
In order to apply our method, let us use a nonphysical seed solution
$u_1(x)$ satisfying equation (\ref{uplus}) for $x_l = -\infty$, i.e.,
let us make in equation (\ref{freegralu}) $B = 0$ and $A = 1$ so that:
\begin{equation} \label{freeuplus}
u_1(x) = e^{kx} .
\end{equation}
With this choice, the calculation of equation (\ref{wuplus}) leads to:
\begin{equation} \label{freew}
w(x) = - \frac{e^{2kx}}{2k} .
\end{equation}
Moreover, the evaluation of equation (\ref{f(x)}) with $x_0 = 0$ produces:
\begin{equation} \label{freef(x)}
f(x) = f_0 + \frac{1-e^{2kx}}{8k^3}.
\end{equation}
Note that this function does not have nodes for
$$
f_0 < - \sigma_- = - \frac{1}{8k^3}.
$$
Hence, it is convenient to reparametrize this domain in the way:
\begin{equation} \label{reparametrize}
f_0 = - \frac{1}{8k^3} - \frac{e^{2kx_1}}{8k^3},
\end{equation}
where $x_1 \in (-\infty,\infty)$. Thus, it is straightforward to show
that:
\begin{equation} \label{freef(x)pt}
f(x) = - \frac{e^{k(x+x_1)}}{4k^3} \cosh[k(x - x_1)].
\end{equation}
Finally, by plugging equations (\ref{freeuplus},\ref{freef(x)pt}) into
equation (\ref{v3b}), the hyperconfluent third-order SUSY partner potential
of the free particle turns out to be:
\begin{equation} \label{v3free}
V_3(x) = - 2 k^2 {\rm sech}^2[k(x - x_1)].
\end{equation}
This is the well know P\"oschl-Teller potential with one bound state at
$\epsilon = E_0 = - k^2$, which has been also derived through first-order SUSY (see
e.g. \cite{ju96}, page 30) and confluent second-order SUSY techniques \cite{fs03}.

\subsection{Coulomb potential}

Working in spherical coordinates, separating the angular ones $\theta,
\phi$, and making $\hbar = e = m = 1$, the three-dimensional stationary
Schr\"odinger equation for the Coulomb potential $-e^2/r$ leads to a
one-dimensional problem characterized by the effective potential
\begin{equation} \label{coulombveff}
V_0(r) = - \frac2r + \frac{\ell(\ell + 1)}{r^2},
\end{equation}
where $0\leq r < \infty$, $\ell = 0,1,\dots$ The discrete energy levels $E_n$
of $H_0$, for a fixed value of $\ell$, take the form $E_n = -
1/(n+\ell + 1)^2, \ n = 0, 1, 2,\dots$ In order to apply our method, let us
employ here the normalized ground state eigenfunction,
\begin{equation} \label{coulombu}
u_1(r) = \frac{1}{(\ell + 1)\sqrt{(2\ell + 1)!}}
\left( \frac{2r}{\ell + 1} \right)^{\ell + 1} e^{- \frac{r}{\ell + 1}} ,
\end{equation}
associated to the eigenvalue $E_0 = - 1/(\ell + 1)^2$. Let us start
by calculating the $w(r)$ of equation (\ref{wuplus}) with $r_l=0$, which leads to:
\begin{equation} \label{coulombw}
w(r) =  - \frac{\gamma(2\ell + 3,\frac{2r}{\ell + 1})}{(2\ell + 2)!} =
e^{-\frac{2r}{\ell + 1}}\sum_{k=0}^{2\ell + 2} \frac{1}{k!}\left( \frac{2r}{\ell + 1}
\right)^k - 1 =
- e^{-\frac{2r}{\ell + 1}}\sum_{k = 2 \ell + 3}^{\infty} \frac{1}{k!}\left( \frac{2r}{\ell + 1}
\right)^k,
\end{equation}
$\gamma(a,x)$ being an incomplete Gamma function. Using this result and the expression
for $u_1(r)$ of equation (\ref{coulombu}) it turns out that:
\begin{equation}
\frac{w(r)}{u_1(r)} = - (\ell + 1) \sqrt{(2 \ell + 1)!} \ \, e^{-\frac{r}{\ell + 1}}
\sum_{k = 2 \ell + 3}^{\infty} \frac{1}{k!}\left( \frac{2r}{\ell + 1}
\right)^{k - \ell - 1},
\end{equation}
which vanishes for $r\rightarrow 0$, as required. The calculation of the $f(r)$ of equation
(\ref{f(x)}) with $r_0 = 0$ produces now:
\begin{eqnarray}
f(r) & = &  f_0 - \frac{(\ell + 1)^3 (2 \ell + 1)!}{2}
\sum_{k = 2 \ell + 3}^{\infty} \sum_{m = 2 \ell + 3}^{\infty}
\frac{\gamma(k + m - 2\ell - 1,\frac{2r}{\ell + 1})}{k! \, m!} \nonumber \\
& = &  f_0 - \frac{\gamma(2\ell + 3,\frac{2r}{\ell + 1})}{2 \, \Gamma(2\ell + 4)}
\, r^2 \, {}_2F_2\left(1,2;3,2\ell + 4;\frac{2r}{\ell + 1}\right) \nonumber \\ && + \frac{(\ell + 1)^2}4
\sum_{m = 0}^{\infty} \frac{\gamma(m + 2\ell + 5,\frac{2r}{\ell + 1})}{(m + 2) \,
(m + 2\ell + 3)!} , \label{coulombf}
\end{eqnarray}
which is nodeless in $(0,\infty)$ for $f_0\leq 0$. The hyperconfluent third-order SUSY partner
of the effective potential (\ref{coulombveff}) becomes finally:
\begin{eqnarray}\label{v3finalco}
V_3(r) = - \frac2r + \frac{(\ell + 1)(\ell + 2)}{r^2} + 2 \left[\frac{w^2(r)}{f(r) u_1^2(r)}\right]',
\end{eqnarray}
where $u_1(r), \ w(r)$ and $f(r)$ are given  by equations (\ref{coulombu},\ref{coulombw}) and
(\ref{coulombf}) respectively.

The first two terms of equation (\ref{v3finalco}) correspond to an effective potential
different from the initial one (compare equation (\ref{coulombveff})). This difference is
also reflected in the energy levels of a potential composed
only of these two terms, which are given by $E_n = - 1/(n+\ell + 1)^2, \ n = 1, 2,\dots$ Thus, it is
natural to interpret that the third term of equation (\ref{v3finalco}) is the main responsible of
supporting the ground state energy of $V_3(r)$ at $E_0 = - 1/(\ell + 1)^2$.

Let us note that the family of hyperconfluent third-order SUSY partner potentials given by
equation (\ref{v3finalco}) is different from the ones which have been derived either by
first-order SUSY \cite{fe84,ad88,jr98,ro98a,ro98b} or by second-order SUSY transformations
\cite{ro98a,ro98b,fs05} (just compare the centrifugal terms of each family).

In particular, for $\ell = 0$ it turns out that, departing from the Coulomb potential without
centrifugal term, $V_0(r) = - 2/r$, we arrive at a new one-dimensional potential with a non-trivial
centrifugal term given by
\begin{eqnarray}\label{coulomvl0}
V_3(r) = - \frac2r + \frac{2}{r^2} + 2 \left[\frac{w^2(r)}{f(r) u_1^2(r)}\right]',
\end{eqnarray}
where now
\begin{eqnarray}
\hskip-0.5cm && u_1(r) = 2r e^{- r}, \label{coulombu0} \qquad w(r) =  - \frac12 \gamma(3,2r) =
(2r^2+2r+1)e^{-2 r}-1 ,  \label{coulombw0} \\
\hskip-0.5cm && f(r) = f_0 - \frac{1}{12}\gamma(3,2r)
\, r^2 \, {}_2F_2\left(1,2;3,4;2r\right) + \frac{1}4
\sum_{m = 0}^{\infty} \frac{\gamma(m + 5,2r)}{(m + 2) \,
(m + 3)!}.  \label{coulombf0}
\end{eqnarray}
As an illustration, the isospectral potentials $V_0(r) = - 2/r$ and
the $V_3(r)$ of equations (\ref{coulomvl0}-\ref{coulombf0}) for $f_0 = - 1/10$
as functions of $r$ are shown in figure 1. The corresponding energy levels of $H_3$
and $H_0$ are given by $E_n = - 1/(n + 1)^2, \ n = 0, 1, 2,\dots$

\begin{figure}[ht]
\centering \includegraphics[width=10cm]{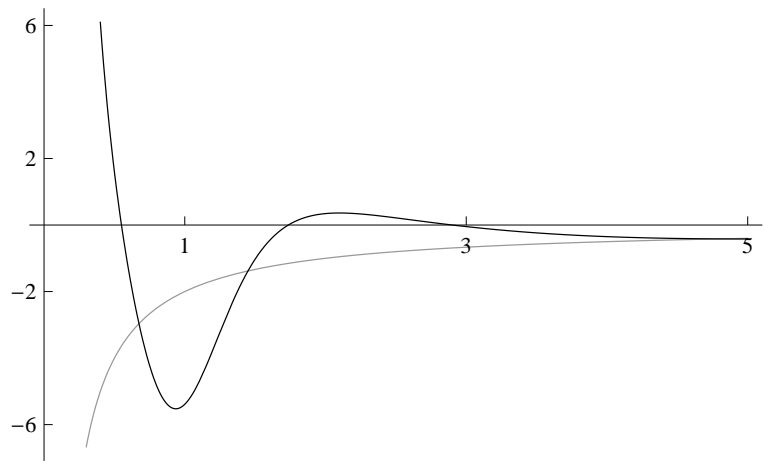}
\caption{Coulomb potential $V_0(r) = - 2/r$ (gray curve) and its
hyperconfluent third-order SUSY partner $V_3(r)$ given by equations
(\ref{coulomvl0}-\ref{coulombf0})
for $f_0 = - 1/10$.}
\end{figure}

Let us note that the well arising in $V_3(r)$ at the neighborhood
of $r \approx 1$ is induced by the third term of equation (\ref{coulomvl0}), which
is also the responsible of supporting the ground state energy at $E_0 = -1$.
Let us recall that this level was not present in the effective potential of
equation (\ref{coulombveff}) with $\ell = 1$.

\section{Conclusions}

In this paper we have addressed the hyperconfluent third-order SUSY QM through two
different (but equivalent) approaches, namely, direct and iterative one. It was found the explicit
expression for the Wronskian, the most relevant quantity which determines the form of the new
potentials, the eigenfunctions of the associated Hamiltonians, etc.

The requirements for the seed solution to produce non-singular SUSY transformations
were as well explicitly determined. Note that, from considerations taking into account
the order of the transformation, through the hyperconfluent third-order SUSY QM one
obtains a three-parametric family of potentials (for a fixed factorization energy).
However, since we had to impose two requirements on the solutions employed
in the iterative approach, it turns out that the non-singular potentials for $r\in(0,\infty)$ belong
just to a one-parametric subset of the general three-parametric family which one is able to
build up.

Our general procedure was illustrated by means of the free particle and the Coulomb
potential. In particular, the last case illustrates clearly that the non-singular one-parametric
family of potentials derived through the hyperconfluent third-order SUSY QM is different either
from the set which can be achieved from a first-order SUSY transformation
\cite{fe84,ad88,jr98,ro98a,ro98b} or from
the one which can be generated through the confluent second-order SUSY transformation \cite{fs05}.

\section*{Acknowledgments}

The authors acknowledge the support of Conacyt. ESH also acknowledges the support of COTEPABE,
EDI and COFAA of IPN, as well as of SNI. The authors wish to thank the comments and suggestions
of the referees of this paper.

\end{document}